\documentclass[10pt,conference]{IEEEtran}

\usepackage{svg}
\usepackage{soul}
\usepackage{cite}
\usepackage{hyperref}
\usepackage{amsmath,amssymb,amsfonts}
\usepackage{graphicx}
\usepackage{textcomp}
\usepackage{xcolor}
\usepackage{caption}
\usepackage{adjustbox}
\usepackage{subcaption}

\usepackage[left=0.680in, right=0.660in, bottom = 1in, top=0.7in]{geometry}
\usepackage{url,amssymb,threeparttable,multirow,booktabs, tabularx}
\def\BibTeX{{\rm B\kern-.05em{\sc i\kern-.025em b}\kern -.08em
    T\kern-.1667em\lower.7ex\hbox{E}\kern-.125emX}}
\usepackage{algorithm}
\usepackage{algpseudocode}
\usepackage{pifont}

\begin{document}

\title{Deep Reinforcement Learning Optimized Intelligent Resource Allocation in Active RIS-Integrated TN-NTN Networks}

\author{
\IEEEauthorblockN{Muhammad Ahmed Mohsin\textsuperscript{1}, Hassan Rizwan\textsuperscript{2}, Muhammad Jazib\textsuperscript{3,4}, Muhammad Iqbal\textsuperscript{4}, Muhammad Bilal\textsuperscript{5}, \\Tabinda Ashraf\textsuperscript{4}, Muhammad Farhan Khan\textsuperscript{6}, Jen-Yi Pan\textsuperscript{4}} 

\IEEEauthorblockA{\textsuperscript{1}Dept. of Electrical Engineering, Stanford University, USA}

\IEEEauthorblockA{\textsuperscript{2}Dept. of Electrical Engineering, University of California, Riverside, USA}

\IEEEauthorblockA{\textsuperscript{3}Dept. of Electrical Engineering, Pakistan Institute of Engineering and Applied Sciences, Pakistan}

\IEEEauthorblockA{\textsuperscript{4}Dept. of Communication Engineering, National Chung Cheng University, Taiwan}

\IEEEauthorblockA{\textsuperscript{5}Dept. of Electrical Engineering, University of California, Irvine, USA}

\IEEEauthorblockA{\textsuperscript{6}School of Computer Science and Information Technology, University College Cork, Ireland}

\IEEEauthorblockA{
Corresponding author emails: \{muahmed@stanford.edu, hrizw002@ucr.edu, iqbal.marjan@gmail.com\}
}
}
\maketitle
\vspace{-40pt}

\begin{abstract}
This work explores the deployment of active reconfigurable intelligent surfaces (A-RIS) in integrated terrestrial and non-terrestrial networks (TN-NTN) while utilizing coordinated multipoint non-orthogonal multiple access (CoMP-NOMA). Our system model incorporates a UAV-assisted RIS in coordination with a terrestrial RIS which aims for signal enhancement. We aim to maximize the sum rate for all users in the network using a custom hybrid proximal policy optimization (H-PPO) algorithm by optimizing the UAV trajectory, base station (BS) power allocation factors, active RIS amplification factor, and phase shift matrix. We integrate edge users into NOMA pairs to achieve diversity gain, further enhancing the overall experience for edge users. Exhaustive comparisons are made with passive RIS-assisted networks to demonstrate the superior efficacy of active RIS in terms of energy efficiency, outage probability, and network sum rate.
\end{abstract}


\section{Introduction}
The proliferation of wireless devices has led to a surge in demand for communication resources. Traditional orthogonal multiple access (OMA) schemes, which allocate dedicated time-frequency resources to each user, are becoming increasingly inefficient. NOMA has emerged as a promising technology that efficiently tackles the problem by leveraging superposition coding and successive interference cancellation (SIC) to enable multiple users to share the same time-frequency resources. This leads to enhanced spectral efficiency of a system \cite{harounabadi2023toward}, \cite{liu2022reconfigurable}. By allocating power to users based on channel conditions, NOMA can improve fairness and accommodate more users within a given spectrum.

In a multi-cell environment, cell-edge users often suffer from severe inter-cell interference as they are located closer to the boundaries of their cells and are more susceptible to interference from neighboring cells. To mitigate this issue, CoMP transmission is utilized to enable coordinated joint detection and joint transmission among multiple base stations, which can significantly improve the performance of cell-edge users \cite{10464446}.

Terrestrial RIS is deployed in densely populated urban areas where a stable user demand persists, while A-RIS serves dynamic environments where user demands fluctuate rapidly \cite{ge2022active}, \cite{10365648}. Moreover, in an urban environment, there will be obstacles for the UAV that create abandoned areas in the model which are a no-fly zone for UAV. Therefore, it is imperative to optimize UAV trajectory to improve the sum rate and ensure UAV safety. Other works utilize the UAV as a BS which offers lesser flexibility than an RIS \cite{9684973}, since RISs can be dynamically reconfigured to adjust their reflecting properties based on changing channel conditions and even direct beams to users for better network capacity.

Typically, UAV communication systems use passive RISs and optimize their phase shift matrix to improve system throughput. However, utilizing an active RIS and optimizing its amplification factor matrix will substantially impact the overall sum rate of the environment as the comparison in \cite{10001687} concludes that active RIS overcomes the multiplicative fading effect found in passive RIS to ultimately, provide a better sum rate gain.

While DRL has been increasingly applied to optimize systems involving RISs and UAVs, existing research primarily focuses on either continuous or discrete action spaces \cite{ammar2024depth}. To achieve optimal performance, a hybrid approach that combines both continuous and discrete action spaces is essential. DRL algorithms that can use both action spaces simultaneously would offer promising solutions for optimization. Furthermore, the optimization of the active RIS amplification matrix using DRL has not yet been explored. 

Keeping in view the literature gap as motivation, the contributions of this paper are threefold. A hybrid DRL agent, H-PPO, is utilized to optimize phase shifts, UAV trajectory, NOMA power allocation factors and base station transmit power to achieve maximum sum rate under optimal SNRs. Phase shifts for terrestrial RIS are optimized to cancel the interference from the non-CoMP base station at the edge user to maximize user experience and network capacity at the edge. Exhaustive results comparisons are made with passive RIS to demonstrate the efficacy of A-RIS in different optimization settings for various performance metrics like outage, sum rate, and fairness.


\section{System Model \& Problem Formulation}

\subsection{System Description}
We consider an active RIS-assisted CoMP-NOMA downlink SISO transmission network with UAV deployment. The network is distributed in $M$ circular grids with a single antenna BS at the center of each grid $m$, where $m \in \mathcal{M} = \{1,2,..., M\}$. Every BS$_m$ in grid $m$ utilizes two-user NOMA to provide a downlink channel to center users and edge users. The center users reside within the radius of the grid while being served by BS$_m$ and are denoted as U$_c^m$, where $c \in \mathcal{C} = \{1,2,..., C\}$ represents all center users. On the other hand, the edge users reside outside the grid radius and are denoted as U$_e^m$ served by BS$_m$, where $e \in \mathcal{E} = \{1,2,..., E\}$ represents all edge users. All users in the system can be denoted as $\mathcal{U} = \mathcal{C} \cup  \mathcal{E}$, while each user $u \in \mathcal{U}$ is treated as either an edge user or a center user. The total UAV flight time is divided into $t$ time slots where $t \in \mathcal{T} = \{1, ..., t_s\}$, where $t_s$ is the total flight time for UAV. Moreover, UAV cannot fly in the forbidden zones present in the system model which are modeled as circular disks of radius $d_{\min}$ centered around different obstacles where an obstacle $o \in \mathcal{O} = \{1, 2, \ldots, O\}$.
    
    The system model includes two active RIS: one fixed on a building equidistant from all grids R$_G$, and the other is located on a mobile UAV R$_U$ which changes its location to provide maximum sum rate to users in the network. Both RISs serve all users in the network while being operated by a microcontroller to alter the phase shifts. Active RIS $R$, where R $=$ \{R$_G$, R$_U$\}, in the model, is equipped with $k$ reflecting elements where $k \in \{1, 2, ..., K\}$. The signals reflected multiple times by the RISs are assumed to exhibit minimal power due to significant path loss as considered by the multi-RIS environment in \cite{10570759}, and this work carries the same assumption. Specifically, $\forall m \in \mathcal{M}, u \in \mathcal{U}$, and $o \in \mathcal{O}$, the positions of BS$_m$, $U_u$, and $O_o$ are represented by $p_m = (x_m, y_m, H_B)$, $p_u = (x_u, y_u, 0)$, and $p_o = (x_o, y_o, H_o)$, respectively, where $H_B$ and $H_O$ are the heights of the BSs and obstacles, respectively. The position of aerial RIS at time slot $t$ is denoted as $p_{R_U}[t] = (x_{R_U}[t], y_{R_U}[t], H_{R_U})$ and it hovers over a specific area $A$. As mentioned earlier, the UAV is mobile and changes its horizontal position in the xy-plane, while maintaining a fixed altitude $H_{R_U}$.


\begin{figure}[t!]
    \centering
    \includegraphics[width=0.75\linewidth]{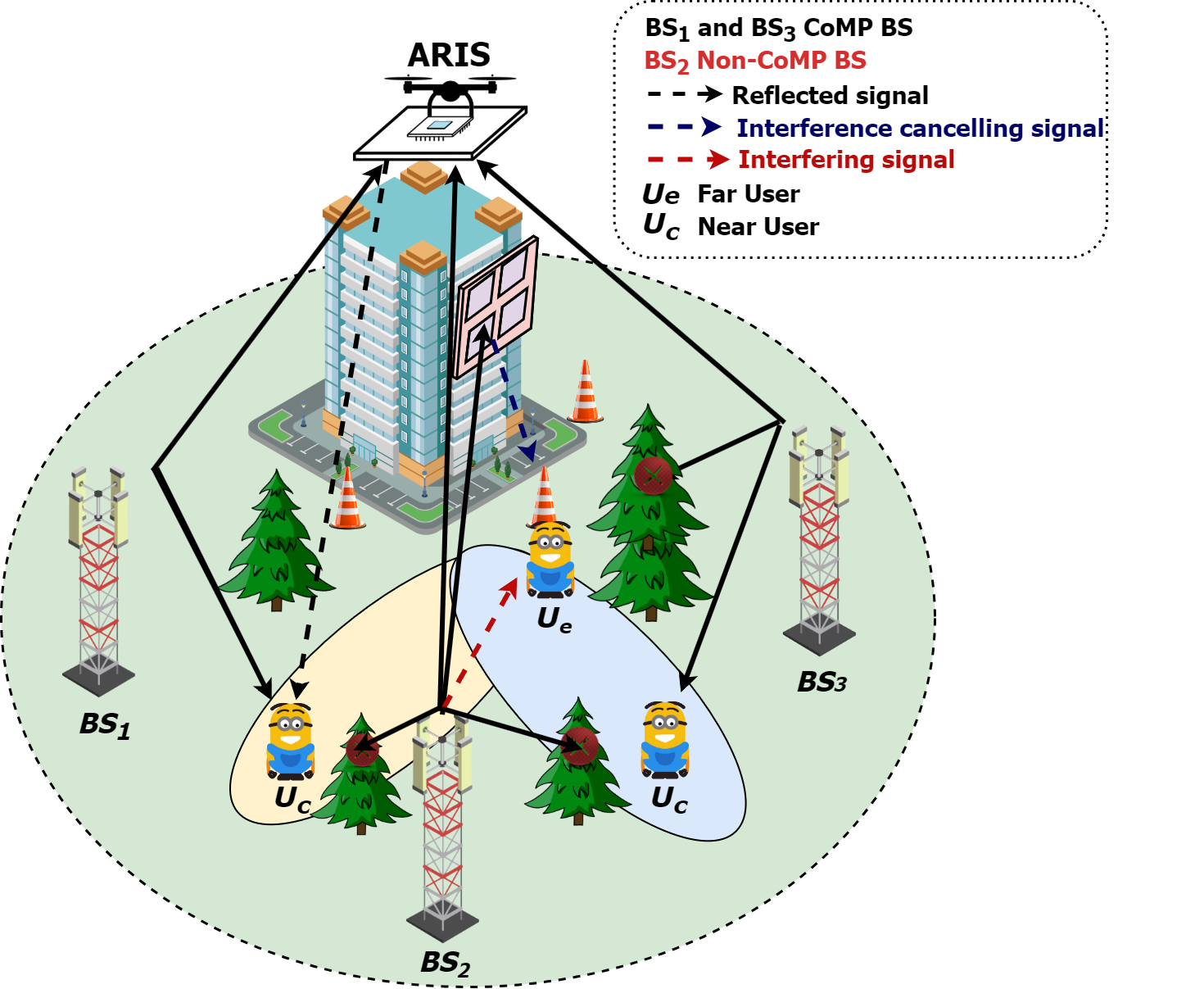}
    \caption{RIS-assisted TN-NTN with coordinated NOMA pairing.}
    \label{fig:enter-label}
\end{figure}

\subsection{Channel Model}
The model includes line-of-sight and non-line-sight paths, so Rayleigh and Rician fading models are used. Due to scattering in the environment, Rayleigh fading channel models are used to simulate channels between the BSs and users. The channel between BS$_m$ and user U$_u$ is denoted as $h_{m,u}$ and expressed for a time interval $t$ as
\begin{equation}
h_{m,u}[t] = \sqrt{\frac{\rho_0}{PL(d_{m,u})}}v_{m,u}[t],
\end{equation}
where $\rho_0$ is a reference path loss at 1 m, $PL(d_{m,u})$ is the large-scale path loss modeled as $PL(d_{m,u}) = (d_{m,u})^{\alpha}$, such that $\alpha$ is the path loss exponent, $d_{m,u}$ is the distance between BS$_m$ and $U_u$, whereas the small-scale Rayleigh fading coefficient is $v_{m,u} \sim \mathcal{R}(1)$. Since a direct line of sight exists between the BS and the RIS on the UAV and RIS on the building, this channel is simulated using Rician fading. The channel between the BS$_m$ and $R$ is denoted as $h_{m,R}$ and expressed for time interval $t$ as 
\begin{equation} 
h_{m,R}[t] = \sqrt{\frac{\rho_0}{PL(d_{m,R})}} \left( \sqrt{\frac{\kappa}{1+\kappa}} g_{m,R}^{\text{LoS}} + \sqrt{\frac{1}{1+\kappa}} g_{m,R}^{\text{NLoS}} \right),
\end{equation}

where $\kappa$ is the Rician factor, $g_{m,R}^{\text{LoS}} \in \mathbb{C}^{K \times 1}$ is the LoS channel vector given by
$$g_{m}^{\text{LoS}} = \left[ 1, \ldots, e^{j(k-1)\pi \sin(\omega)}, \ldots, e^{j(K-1)\pi \sin(\omega)} \right]^T,$$
where $\omega$ represents the angle of arrival of the LoS component at $R$ while ${g}_{m,R}^{\text{NLoS}} \in \mathbb{C}^{K \times 1}$ is the NLoS component which follows Rayleigh fading as previously described \cite{10464446}. This work assumes perfect channel state information (CSI) to avoid unnecessary overheads.


\subsection{Active RIS Configurations}


The active RIS in the network amplifies the incoming signal, resulting from the reflection-type amplifiers employed by the active RIS elements backed by a power supply. The outgoing signal from the active RIS with $K$ reflecting elements is expressed as:
\begin{equation}
y[t] = \underbrace{P[t] \Theta[t] x[t]}_{\text{Desired signal}} +   
\underbrace{P[t] \Theta[t] v}_{\text{Dynamic noise}} 
+ \underbrace{n_s}_{\text{Static noise}} ,
\end{equation}
where $P[t] = \operatorname{diag}(p_1[t], \dots, p_K[t]) \in R^{K \times K}$ represents the amplification matrix of the active RIS for a time slot $t$. Each active RIS element can be denoted as $1 \leq p_N \leq \mathcal{S}$ where $\mathcal{S}$ is the maximum amplification an element can provide. Further, $x[t] \in \mathbb{C}^K$ denotes the incoming RIS signal, $y[t] \in \mathbb{C}^K$ denotes the outgoing RIS signal and $\Theta[t] = \mathrm{diag} \left( a_1 e^{j \theta_1[t]}, a_2 e^{j \theta_2[t]}, \ldots, a_k e^{j \theta_K[t]} \right) \in \mathbb{C}^{K \times K}$ for a time instance $t$, where $a_{k} \in (0,1]$ denotes the amplitude coefficient and $\theta_{k} \in [-\pi, \pi])$ denotes the phase shift of the $k$-th RIS element. Also, $diag(.)$ is the diagonalization operation that is performed on the RIS phase shift matrix. 
    Active RIS elements consume additional power to amplify the reflected signals while generating thermal noise divided into two components: dynamic and static. Here, we assume only the dynamic noise and ignore static noise, since the study in \cite{10001687} shows that static noise is disproportionately lesser in magnitude than dynamic noise and can be neglected. The dynamic noise variable is denoted as $v$ and modeled as $v \sim \mathcal{CN}(0_K, \sigma^2 I_K)$, where $\mathcal{CN}(\mu, \Sigma)$ signify the complex multivariate Gaussian distribution with mean and variance as $\mu$, $\Sigma$, respectively. The parameters $I_K$ and $0_K$ denote a $K\times K$ identity matrix and a $K\times1$ zero vector, respectively.

\subsection{NOMA Model}
All BSs serve a centre user and an edge user, so by following the 2-user NOMA tradition the power of the BS is distributed among the two users where the edge user receives a higher power signal \cite{9681835}, \cite{9259258}. The signal transmitted by the BS$_m$ can be expressed as $x_{m}[t]=\sqrt{(1-\lambda_{m})p_{t}}x_{m,c}[t]+\sqrt{\lambda_{m}p_{t}}x_{m,e}[t],$ where $x_{m,c}[t]$ and $x_{m,e}[t]$ is the received signal for U$_c$ and U$_e$. All BSs in the system have the same transmit power, $p_t$, and the power allocation factor of U$_e$ is denoted as $\lambda_m$. The power allocation factor $\lambda$ should satisfy the constraint $0.5 < \lambda < 1$, since the edge user is further from the BS and demands higher power allocation \cite{9259258}. 

Each U$_c$ receives a direct signal from the BS, an indirect reflected link from the RIS, interference from another BS serving a user of another cell, and inherent noise of active RIS. Ultimately, the signal received by U$_c$ is expressed as
\begin{equation}
\begin{aligned}
    y_c[t] &= \underbrace{H_{m,c}[t] p_t x_m[t]}_{\text{desired signal}} + \underbrace{\sum_{j \neq m}^{M} H_{j,c}[t] p_t x_j[t]}_{\text{inter-user interference}} \\
    &\quad + \underbrace{{h}_{R,c}[t]{P[t]} {\Theta[t]} {v}}_{\text{active RIS noise}} {+n_u},
\end{aligned}
\end{equation}
where $n_u$ denotes the additive white Gaussian noise for user $u$ as $n_u \sim \mathcal{N}(0, \sigma^2)$ and the channel from BS$_m$ to the user U$_c$ is defined as $H_{m,c}[t] = h_{m,c}[t] + {h}_{R,c}[t]{P[t]}{\Theta[t]}{h}_{m,R}$, which includes the direct link and the indirect link involving RIS. The inter-user interference is ICI obtained when U$_c$ experiences a signal broadcast from BS$_j$ \cite{10185599}. After utilizing SIC, U$_c$ decodes the signal for U$_e$ and removes it from the signal received to decode its own signal. Therefore, the achievable rate for U$_c$ to decode the signal for U$_e$ is as follows
\begin{equation}
\mathcal{R}_{c \rightarrow e}[t] = \log \left( 1 + \frac{\lambda_{m} \gamma_{m}}{(1 - \lambda_{m}) \gamma_{m} + 1} \right),
\end{equation}
where $\gamma_{m} = \frac{p_t |h_{m,c}[t]|^2}{\Psi} \text{ and } \Psi = p_t|h_{j,c}[t]|^2 + \sigma^2.$ Whereas, the data rate for U$_c$ to decode its own signal is  as follows
\begin{equation}
\mathcal{R}_{c}[t] = \log \left( 1 + (1 - \lambda_{m}) \gamma_{m} \right). \label{eq:Rc}
\end{equation}

The signal received by U$_e$ is received via BS$_m$ and RIS $R$ can be expressed as 
\begin{equation}
\begin{aligned}
y_e &= (h_{m,e}[t] + h_{R,e}^H[t] P[t] \Theta[t] h_{m,R}[t]) x_{m}[t] p_t \\
&\quad + (h_{j,e}[t] + h_{R,e}^H[t] P[t] \Theta[t] h_{j,R}[t]) x_{j}[t] p_t + n_u, 
\end{aligned}
\end{equation}

where $h_{R,e}^H$ represents the Hermitian transpose of the channel between RIS $R$ and user $e$.
The data rate for U$_e$ can be expressed as
\begin{equation}
\mathcal{R}_{e}[t] = \log \left( 1 + \frac{\lambda_{m} \gamma_{m} + \lambda_{j} \gamma_{j}}{(1 - \lambda_{m}) \gamma_{j} + (1 - \lambda_{j}) \gamma_{j} + 1} \right) \label{eq:Re}
\end{equation}

After obtaining sum rate expression for centre user in \eqref{eq:Rc} and for edge user in \eqref{eq:Re}, we can obtain sum rate expression for whole system as 
\begin{equation}
R_{\text{total}}[t] = \sum_{c \in \mathcal{C}} R_{c}[t] + \sum_{e \in \mathcal{E}} R_{e}[t]. \label{eq:Rt}
\end{equation}



\subsection{Energy efficiency}
The energy efficiency of our proposed active RIS-assisted NOMA-CoMP system highlights the trade-off between network performance and energy costs, as follows:
\begin{equation}
\eta_{E} = \frac{R_{\text{total}}[t]}{p_{\text{total}}}.
\end{equation}


The total power consumed in the network comprises power from BS, active RIS, and UAV. 
The total sum rate of the system was expressed in $(\ref{eq:Rt})$, we can express the total power consumed in the network as follows
\begin{equation}
\resizebox{0.5\textwidth}{!}{$
    \begin{split}
        p_{\text{total}} = \underbrace{p_{\text{BS}}}_{\text{BS power}} + \underbrace{\sum_{i=1}^{2} \sum_{k=1}^{K} \left( \eta_{R_i} |P_{{R_i}k} \Theta_{{R_i}k} x_{{R_i}k}|^2 + P_{\text{c}, {R_i}k} \right)}_{\text{RIS power}}
        + \underbrace{P_{\text{UAV}}}_{\text{UAV power}} 
    \end{split}
$},
\end{equation}
where \(p_{BS}\) represents the static power consumption and transmission power at the base station. The RIS power consumption includes
\(\eta_{{R_i}}\) as the amplification efficiency for the RIS ${R_i}$, \(P_{{R_i}k}\) as the incoming signal power for the \(k\)-th element of the RIS $R_i$, \(\Theta_{{R_i}k}\) as the phase shift applied by the \(k\)-th element of RIS ${R_i}$ panel, \(x_{{R_i}k}\) as the incoming signal for the \(k\)-th element, and \(P_{c,{R_i}k}\) denotes the circuit power consumption for the \(k\)-th element of RIS ${R_i}$. Lastly, \(P_{UAV}\) indicates the power consumed by the UAV for its hovering and motion. 

\subsection{Problem Formulation}
In this work, our primary objective is to maximize the sum rate achieved over $t$ time slots. To achieve this goal, we jointly optimize three key control variables: the UAV trajectory denoted as $\mathbf{\widehat{T}} \triangleq \{p_{R_U}[t], \forall t\},$ the RIS phase shifts represented by $\mathbf{\Theta} \triangleq \{\Theta[t], \forall t\},$ the power allocation factors denoted as 
$\mathbf{\Lambda} \triangleq \{\lambda_m, \forall m\},$ and the RIS amplification matrix $\mathbf{P} \triangleq \{p_k[t], \forall k\}$ 
The problem can be mathematically formulated as

\begin{equation}
\max_{\mathbf{P}, \mathbf{\Theta}, \mathbf{\Lambda}, \mathbf{\widehat{T}}} \sum_{t \in \mathcal{T}} R_{\text{sum}}[t] ,
\tag{10a}
\end{equation}
subject to:
\begin{align}
x_R[t], y_R[t] &\in A, \quad \forall t \in \mathcal{T}, \tag{10b} \\
\|\mathbf{p}_R[t] - \mathbf{p}_o\| &\geq d_{\min}, \quad \forall o \in \mathcal{O}, t \in \mathcal{T}, \tag{10c} \\
\theta_k[t] &\in [-\pi, \pi), \quad \forall k \in K, t \in \mathcal{T}, \tag{10d} \\
R_{c}[t] &\geq R_{c}^{\min}, \quad \forall m \in M, t \in \mathcal{T}, \tag{10e} \\
R_e[t] &\geq R_e^{\min}, \quad \forall e \in \mathcal{E}, t \in \mathcal{T}, \tag{10f} \\
\lambda_m &\in (0.5, 1), \quad \forall m \in M, t \in \mathcal{T}, \tag{10g} \\
p_k[t] &\in [1, \mathcal{S}], \quad \forall k \in K, t \in \mathcal{T},  \tag{10h}
\end{align}


\section{Deep Reinforcement Learning-based Proposed Solution}

\subsection{MDP Formulation}
The MDP is defined via the tuple $(\mathbb{S, A, P,} \mathcal{r}, \mathbb{D} ),$ where $\mathbb{S}$ and $\mathbb{A}$ represent the state space and action space, respectively. $\mathcal{r}$ is the reward to the agent for its actions, and $\mathbb{D}$ is the discount factor to balance weights of immediate and future rewards. Lastly, $\mathbb{P}$ denotes the state transition probability i.e., the likelihood of transitioning from one state to another for an action. At each time slot $t$, the agent observes the current state $s_t$, selects an action $a_t$ based on its policy, and transitions to a new state $s_{t+1}$, to get rewarded.

\subsection{State Space}
The state space in a time slot $t$ is composed of the current position of UAV, BS power allocation factor, amplification matrix of active RIS, and achievable sum rates, which can be denoted as $\varrho[t]$, $\zeta$, $\vartheta[t]$, and $\mathbf{R}[t] = \{\mathcal{R}_c[t], \mathcal{R}_e[t], \forall c, e\}$, respectively. Eventually, the state space becomes

\begin{equation}
s_t = \{\varrho[t], \zeta, \vartheta[t], \mathbf{R}[t]\} .
\end{equation}

The dimension for the state space can be expressed as dim($\mathbb{S}_t$) $= 2 + M + K^2 + 2$.
\subsection{Action Space}
The action space for the MDP is composed of the UAV's horizontal movement in xy-plane, the amplification factor and phase shifts of the active RIS, and the power allocation factor of the BSs. Specifically, the action space at time slot $t$ contains the UAV actions $a_R[t] \in \{(-1,0), (1, 0), (0, -1), (0, 1), (0,0)\}$, representing left, right, down, up, and hover action, respectively. The phase shifts $a_{\theta}[t] =
\{\theta_k[t], \forall k\}$, the power allocation factors $a_{\Lambda} = \{\lambda_m, \forall m\}$, and the active RIS amplification matrix $a_{p}[t]=\{p_{k},\forall k\}$.
Thus, the action space can be expressed as
\begin{equation}
a_t = \{a_{R}[t], a_{\theta}[t], a_{\Lambda}, a_{p}[t]\} 
\end{equation}
The dimension for the action space is expressed as dim($a_t$) $= 2 + K + M + K^2$.

\subsection{Reward Function}
The reward function ensures the maximum sum rate while ensuring UAV safety and meeting QoS requirements by penalizing constraint violations when UAV goes OOB (out of bounds). The reward function is defined as

\begin{equation}
R(s_t, a_t) = R_{\text{sum}}[t] + \xi_{\text{dist}} \left( \frac{C}{d_{\text{R$_U$}, \text{U}}[t]} \right)\zeta  - \xi_{\text{OOB}} \cdot \mathbb{I}(\text{OOB}),
\end{equation}

where $\zeta[t] = \mathbb{I}(d_{\text{R$_U$}, \text{U}}[t] < \text{threshold})$ acts as the indicator function for the distance incentive to keep the UAV close to users. Also, $\xi_{\text{OOB}}$ represents the penalty given to the agent when R$_U$ goes out of bounds of the grid which is pointed by the indicator function $\mathbb{I}(\text{OOB})$. Lastly, $\xi_{\text{dist}}$ and $C$ are kept as constants in the expression.



\begin{algorithm}[t!]
\small
\caption{H-PPO Active RIS-Based Energy Optimization}
\begin{algorithmic}[1]
    \State \textbf{Initialize:} Parameters \(\alpha_d, \alpha_c, \zeta, \Psi\)
    \For{iteration \(l = 1, \ldots, L\)}
        \State Receive initial state \(s_0\)
        \For{time step \(t = 0, \ldots, T\)}
            \State Choose continuous actions \(a_p\) and \(a_c\) using \(\pi_{\alpha_c}(s_t)\)
            \For{each RIS element \(n\)}
                \State Energy optimization \(e_n = g(\psi_n, a_p, a_c)\)
                \State Update the gain for the \(k\)-th RIS element based on 
                \Statex\hspace{1.2cm}channel conditions
            \EndFor
            \State Execute actions \(a_t = \{a_R, \Psi, a_p, a_c\}\)
            \State Observe reward \(r_t\) based on achieved sum rate and outage 
            \Statex\hspace{1.0cm}probability
            \State Store the transition \((s_t, a_t, r_t, s_{t+1})\) in experience replay 
            \Statex\hspace{1.0cm}buffer
            \State Compute the advantage function estimate $\hat{A}_{t}$
        \EndFor
    \EndFor
    \For{epoch \(m = 1, \ldots, K\)}
        \State Sample mini-batch experiences \(E\) from the replay buffer
        \State Objective computation using clipped functions for both 
        \Statex\hspace{0.5cm}discrete and continuous actions: $L_{d}^{\text{CLIP}}(\theta_d)$ and $L_{c}^{\text{CLIP}}(\theta_c)$
        \State Optimize policy parameters \(\alpha_d\) and \(\alpha_c\) 
    \EndFor
    \State Obtain previous policy parameters \(\alpha_d^{\text{old}} \leftarrow \alpha_d\) and \(\alpha_c^{\text{old}} \leftarrow \alpha_c\)
    \State Clear
\end{algorithmic}
\end{algorithm}

\subsection{H-PPO algorithm}
We employ a hybrid Proximal Policy Optimization (H-PPO) algorithm to cater to the hybrid action space in our network. The algorithm offers actions in both discrete $a_{R}[t]$ and continuous action spaces $a_{\theta}[t], a_{\Lambda}, a_{p}[t].$ The value function $V(\mathbb{S}_t)$ is used to obtain a variance-reduced advantage function estimate $\hat{A}_t$ to optimize policy. Following the implementation details used in \cite{pmlr-v48-mniha16}, the policy is executed for $T$ time steps, and $\hat{A}_t$ is computed as

\begin{equation}
\hat{A}_{t} = \sum_{k=0}^{\bar{T}-1} \mathbb{D}^k r_{t+k} + \mathbb{D}^{\bar{T}} \hat{V}(s_{t+\bar{T}}) - V(s_t), 
\end{equation}

where $\bar{T}$ is much smaller than the length of the episode $T$. Stochastic policy for discrete and continuous action is generated in the same way so only discrete is mentioned here. To generate the stochastic policy $\pi_{\theta_d}(a_{t}|s_{t})$ for the discrete actions, the corresponding actor network outputs $|\mathbf{a}_R|$ logits, which are then passed through a softmax function to obtain a probability distribution over the available discrete actions. Conversely, the continuous actor network generates the continuous actions $a_{\Phi}$ and $a_{\Lambda}$ by sampling from Gaussian distributions parameterized by the mean and standard deviation
outputs of the network, as dictated by the stochastic policy $\pi_{\theta_c}(a_{t}|s_{t})$. Both $\pi_{\theta_d}(a_{t}|s_{t})$ and $\pi_{\theta_c}(a_{t}|s_{t})$ are optimized independently using their respective clipped surrogate objective functions. For the discrete actions, the objective function is given by
\begin{equation}
L_d^{\text{CLIP}}(\theta_d) = \hat{\mathbb{E}}_t \left[ \min \left( r_t^d(\theta_d) \hat{A}_t, \aleph(r_t^d, \theta_d, \epsilon) \hat{A}_t \right) \right]
\end{equation}
where \(\aleph(r_t^d, \theta_d, \epsilon) = \text{clip}(r_t^d(\theta_d), 1 - \epsilon, 1 + \epsilon)\), \(r_t^d(\theta_d) = \frac{\pi_{\theta_d}(a_t | s_t)}{\pi_{\theta_d^{\text{old}}}(a_t | s_t)}\) is the importance sampling ratio, and \(\epsilon\) is the clipping parameter. 

The optimization objectives of both policies remain decoupled i.e., $\pi_{\theta_d}(a_{t}|s_{t})$ and $\pi_{\theta_c}(a_{t}|s_{t})$ are treated as independent distributions during policy optimization, rather than a joint distribution encompassing both action spaces. The H-PPO algorithm is summarized in Algorithm 1.

\section{Simulation Results}

\subsection{Simulation setup}
In order  to assess the effectiveness of the proposed active RIS framework in a CoMP-NOMA network, we create a virtual representation of an urban setting which includes \( C = 3 \) BSs, \( K = 3 \) users, and \( L = 0 \) obstacles. The initial position of the UAV is set at \( (-5, 0, 40) \, \text{m} \), while the BSs are positioned at coordinates \([ -30, 30, 20 ]\), \([ 30, 30, 20 ]\), and \([ 20, -30, 20 ]\), respectively. Each base station, however, is considered to operate at a unique power level, denoted as \( P_{d,x} \) for each base station \( x \), where \( x \in \{1, 2, 3\} \). The remaining utilities, users inclusive, are distributed randomly throughout the environment.  The network operates at a carrier frequency of \( f_c = 2.4 \, \text{GHz} \), utilizing a bandwidth of \( B = 10 \, \text{MHz} \). The noise power is defined as \(\sigma^2 = -174 + 10 \log_{10}(B), \, \text{dBm} \).
To account for attenuation effects, path loss exponents are defined as \( \beta_n = 2.2 \), \( \beta_o = 3.0 \), \( \beta_k = 3.3 \), and \( \beta_s = 3.7 \). The results presented are averaged over 1000 independent realizations of user positions to ensure reliable performance metrics.

\subsection{Numerical results}
\begin{figure*}[t!]
     \centering
    \begin{subfigure}[t]{0.66\columnwidth} 
         \centering
         \includegraphics[width=\textwidth]{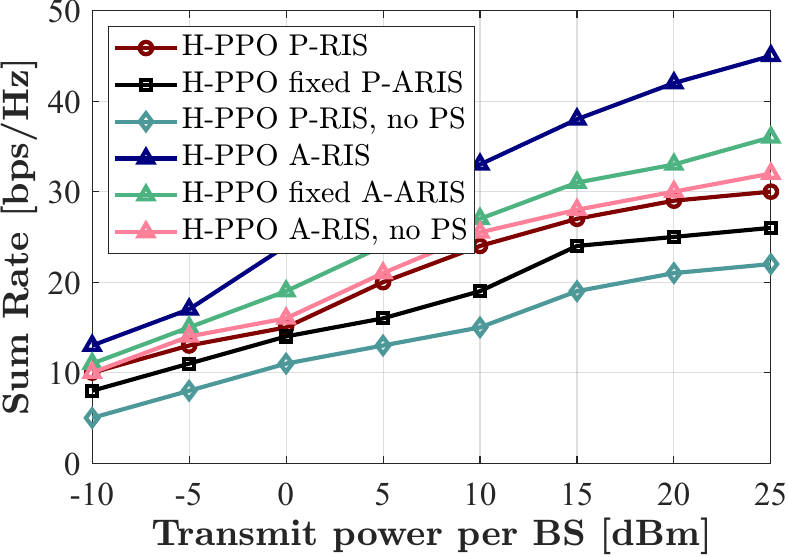}
         \caption{Sum Rate  vs transmit power.}
         \label{fig:Sumrate}
     \end{subfigure}
     \hspace{-0.01\columnwidth} 
     \begin{subfigure}[t]{0.66\columnwidth} 
         \centering
         \includegraphics[width=\textwidth]{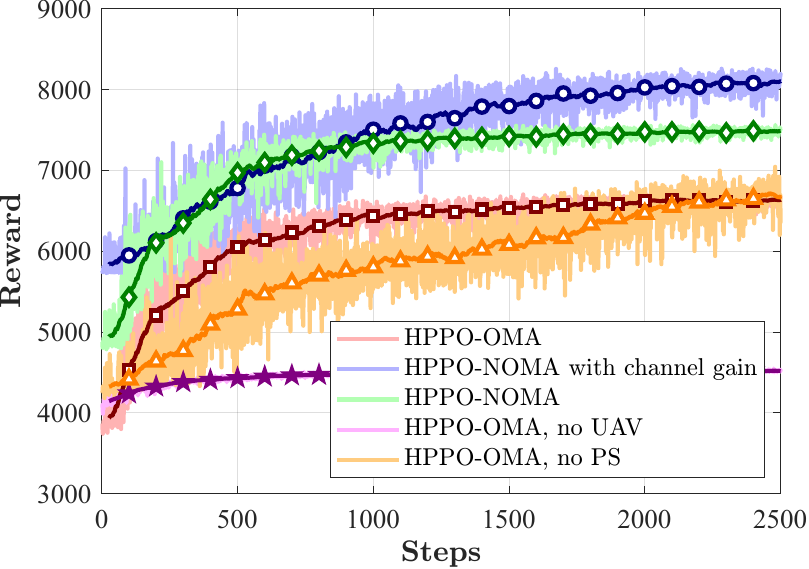}
         \caption{Reward function vs time steps  with normalized RL policies}
         \label{fig:reward}
     \end{subfigure}
     \hspace{-0.01\columnwidth} 
     \begin{subfigure}[t]{0.66\columnwidth} 
         \centering
         \includegraphics[width=\textwidth]{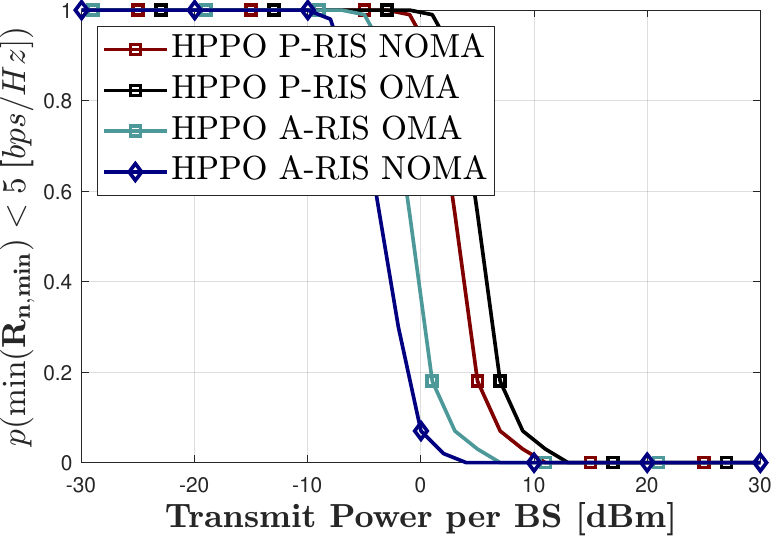} 
         \caption{Outage probability vs. transmit power.}
         \label{fig:outage}
     \end{subfigure}
      \caption{\textbf{(a)} Sum rate performance for baseline methods  [OMA, Brute Force and RL-NOMA] varying transmit power, \textbf{(b)} Training convergence of reward functions for various RL configurations over timesteps, \textbf{(c)} Outage probability comparison against transmit power per base station for NOMA and OMA base.}
     \label{fig:2}
\end{figure*}

 In Fig.\ref{fig:Sumrate} investigation of the relationship between transmit power $p_t$, where $p_t \in [-30 \, \text{dBm},  30 \,  \text{dBm}]$  and sum rate  $R_k$ is made, comparative analysis between different H-PPO-RIS configuration were taken in to account, highlighting active RIS improvement. Compared with H-PPO P-RIS $R_k$, H-PPO A-RIS achieves $(7 \, \text{bps/Hz})$ at $p_t$ of $-10 \, \text{dBm}$, performing better than  H-PPO P-RIS configuration. The cumulative rate $R_k(T, P_{\mathrm{tx}},  h, \gamma)$  for H-PPO P-RIS and H-PPO A-RIS increases to  $(15 \,  \text{bps/Hz})$  and  $(20 \, \text{bps/Hz})$,  respectively,  when $p_t$ reaches $25 \, \text{dBm}$.  This represents a 33\% improvement in sum rate, emphasizing the amplification capability of active RIS enabled by NOMA, which outperforms passive configurations..
 
 Our investigation further spread to the learning algorithm efficacy for energy optimization. Fig.\ref{fig:reward} illustrates the progression of reward dynamics $r$ across multiple training steps, with HPPO-NOMA setup displaying a steeper growth, from $0.2$ to $1.0$, showing higher learning efficiency, indicating NOMA's improved energy management.

Also, we have evaluated the outage probabilities of our configurations, In Fig.\ref{fig:outage} at $p_t$ of $-30 \, \text{dBm}$ to $-10 \, \text{dBm}$,  outage probability of all configuration remains same for all worst users rate. The HPPO A-RIS NOMA configuration has the least outage probability $0.20$ at $30 \, \text{dBm}$, indicating the worst user rate could increase from $5 \, \text{bps/Hz}$, hence outperforms all other configurations in consideration.

\begin{figure*}[t!]
     \centering
     \begin{subfigure}[t]{0.66\columnwidth} 
         \centering
         \includegraphics[width=\textwidth]{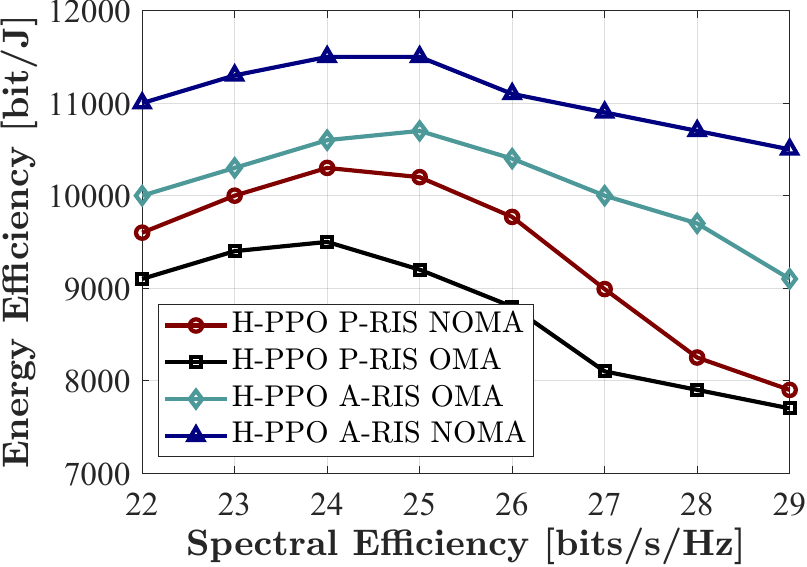}
         \caption{Energy efficiency vs. spectral efficiency.}
         \label{fig:Energy effeicency}
     \end{subfigure}
     \hspace{-0.01\columnwidth} 
     \begin{subfigure}[t]{0.66\columnwidth} 
         \centering
         \includegraphics[width=\textwidth]{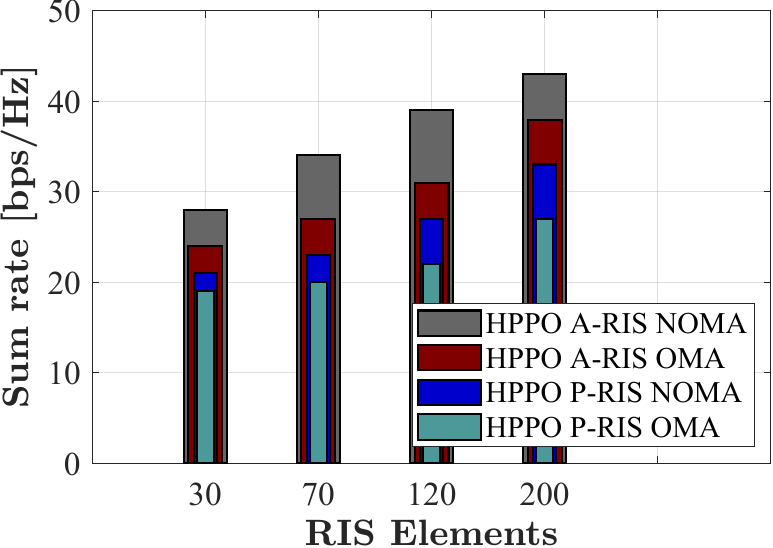}
         \caption{Sum rate vs. RIS elements}
         \label{fig:ELEMENTS}
     \end{subfigure}
     \hspace{-0.01\columnwidth} 
     \begin{subfigure}[t]{0.66\columnwidth} 
         \centering
         \includegraphics[width=\textwidth]{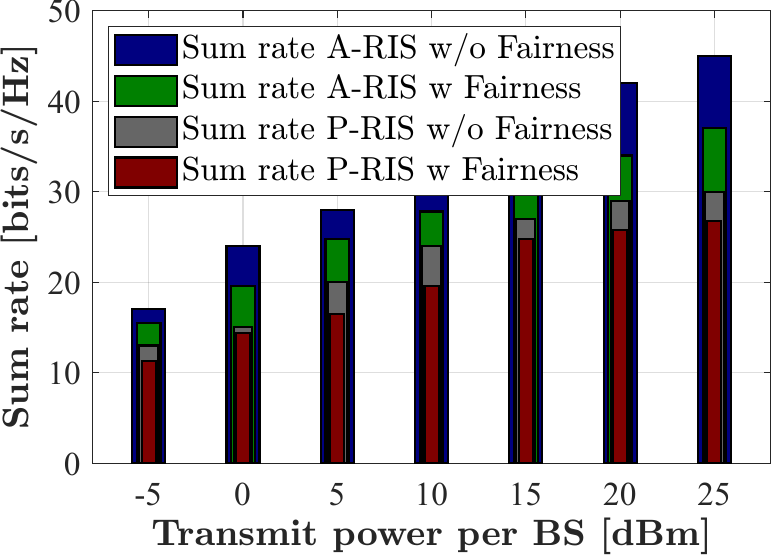} 
         \caption{Sum rate vs. transmit power}
         \label{fig:fairness}
     \end{subfigure}
     \caption{\textbf{(a)} Energy efficiency comparison against spectral efficiency between  different RIS configuration , \textbf{(b)} Sum rate comparison with RIS elements n belong to [30,70,120,200] for different RIS configuration \textbf{(c)} Sum rate comparison against baseline methods [DRL NOMA (w and w/o fairness), OMA].}
     \label{fig:3}
\end{figure*}

Fig.\ref{fig:Energy effeicency} illustrates the relationship between energy efficiency and spectral efficiency for various configurations. The H-PPO A-RIS NOMA configuration demonstrates the highest performance, reaching a peak energy efficiency of $11,500 \, \text{bit/J}$ at $24 \, \text{bits/s/Hz}$.This improvement results from the active RIS's ability to dynamically adjust reflections, enhanced further by the NOMA strategy, achieving a 20\% improvement over passive RIS configurations.
Fig.\ref{fig:ELEMENTS} shows the sum rate for different configurations as the number of RIS elements increases. $K$ increases from 30 to 200 for all configurations indicating that adding more RIS elements $K$ enhances the system's ability to optimize reflections and HPPO A-RIS NOMA consistently shows the highest $R_k$ across all $K$ highlighting the advantage of active RIS with NOMA.

Lastly, we have Fig.\ref{fig:fairness} illustrating the relationship between $p_t$ per base station and $R_k$ for different RIS configurations with fairness and without fairness, indicating the trade-off between system capacity and fairness among users. As the $p_t$ increases from $-5 \, \text{dBm}$ to $25 \, \text{dBm}$, the $R_k$ improves for all configurations. With an even power distribution among center and edge users, fairness slightly reduces the $R_k$. H-PPO A-RIS-NOMA with fairness achieves the highest performance among all other configurations.

\section{Conclusion}
This paper proposes a DRL-based resource allocation framework for active RIS-assisted TN-NTN networks using CoMP-NOMA.The active RIS response system incorporating COMP-NOMA significantly achieved higher sum rates and energy efficiency than the traditional passive RIS and other OMA configurations. These results show that the active RIS integration with CoMP-NOMA and DRL is an efficient option for resource allocation and better system performance. Future work could explore the integration of energy harvesting technologies with active RIS, or using mmWave and terahertz communication to open exciting research avenues.
\section{acknowledgements}
This work was partly supported by the National Science and Technology Council, Taiwan, R.O.C., under Grants  NSTC 113-2622-E-194 -011  and  NSTC 113-2218-E-194 -003. This work was also partially supported by the Advanced Institute of Manufacturing with High-tech Innovations (AIM-HI) from The Featured Areas Research Center Program within the framework of the Higher Education Sprout Project by the Ministry of Education (MOE) in Taiwan.

\footnotesize
\bibliographystyle{IEEEtran}
\bibliography{main}
\end{document}